# Game-Based Approach for QoS Provisioning and Interference Management in Heterogeneous Networks

A.S.M. Zadid Shifat[1], Mostafa Zaman Chowdhury[1,2], and Yeong Min Jang[2,*]
[1]Dept. of Electrical and Electronic Engineering, Khulna University of Engineering and Technology (KUET), Khulna-9203, Bangladesh
[2]Dept. of Electronics Engineering, Kookmin University, Seoul-02707, Korea
E-mail: Zadid_shifat@outlook.com, mzceee@yahoo.com, *yjang@kookmin.ac.kr
*Corresponding author

*Abstract*— For the current fifth-generation era, the deployment of small cells in residential and commercial areas plays an imperious preamble in improving network coverage and the quality of service (QoS). Major technical problems associated with the mass deployment of small cells such as femtocells are interference management and QoS provisioning. These are important for service-providing operators because the system capacity and achievable data rates mainly depend on interference. Future generation wireless networks will use autonomous and distributed architecture for ameliorating the efficacy and flexibility of communication systems. In this paper, we propose a game theory based model along with dynamic channel allocation and self-optimizing power control scheme for resolving priority-based access exposure by applying the concept of primary and secondary users. It is expected that the consumers will experience better QoS with reduced interference levels, and the service-providing operators will be able to increase their revenue while ensuring optimal price for the consumers. We assimilated extensive numerical results to demonstrate the efficacy of our proposed model.

## I. Introduction

With increasing technological advancement, there has been a continual increase in the number of active wireless terminal nodes. As a result, wireless network service providers have been receiving demands for higher data rates and better quality of service (QoS). The modern fifth-generation (5G) mobile networks are expected to provide high throughputs for a wide variety of services with different QoS requirements; they are also expected to provide tactile internet with minimum end-to-end latency [1], [2]. The next generation communication networks need to improve the indoor coverage and capacity and bounce back high-data-rate services to the users because the number of mobile phone users is increasing instantaneously. In order to meet the needs of the increasing traffic demands and to improve QoS as well as reduce costs, energy efficiency has become one of the major design objectives for next generation network systems [3]. This is because the measure of delay is more relevant to distinguish user experience than the minimal data rate requirement [4]. Small cells such as femtocells help overcome these hindrances; they have low power consumption features, and they increase the capacity of wireless networks and extend cellular coverage [5], [6]. Nevertheless, no authors have suggested proper frequency band allocation for indoor femtocell. It may happen that, femtocell is using the same frequency band of macrocellular base stations (MBSs) and causing adequate interference.

Femtocell provides some economic benefits compared to the traditional cell-partitioning approach, which would require a large number of expensive base stations (BSs). By deploying femto base stations (FBSs), indoor users are able to enjoy high-speed data rates and other communication services because of the proximity between the BSs and FBSs. Approximately 90% of data services and 60% of phone calls take place in indoor environments. Another survey illustrated that approximately 30% of businesses and 45% of residential users experience poor indoor coverage [7]. The traffic loads of MBSs can be determined by their corresponding femto access points (FAPs) by installing femtocells [8]. It is helpful to delimit the cross-tier interference between the femtocell user equipment (FUE) and MBS because the femtocells operating in the licensed spectrum are based on the MBSs [9]. Although it is beneficial to implement femtocells for capacity improvement, traffic load management, and proper frequency allocation with reduced interference remains a major concern. Moreover, femtocells and macrocells share the same spectrum, which leads to a deterioration in each other's performance; therefore, cross-polarization allocation concept was started in the line-of-sight systems as a way of sequestrating the interfering signal from the desired signal [10]. Despite being capable to reduce co-tier interference, cross polarization concept does not exhibit any satisfactory outcome in case of cross-tier interference. Researchers are endeavoring to develop a priority-based self-explanatory interference management scheme such that the QoS requirements of the macrocell users with higher access priority are always considered, and the network capacity is assimilated by the femtocell users so that their own performances are optimized [11], [12]. Nevertheless, a major problem in this scheme is the assurance of accessibility in a systematic way along with different access tariffs applicable for these two types of users. A cost-effective frequency planning strategy along with dynamic channel allocation is proposed in [13] with a view to dealing with the violations of QoS, which would ensure the reduction of interference. Certain researchers have presented a singular solution [14], [15] whenever the minimum signal-to-noise-plus-interference ratios (SINRs) of all the users can be corroborated. Despite being outstanding concepts, both of them will face great challenges if there is no dynamic power control and priority based access scheme is adopted. Efficient call admission control methodology along with user removal algorithm schemes have been used for infeasible SINR targets in [16]. Here, the authors have considered coordinated multipoint transmission along with the joint transmission scheme for cellular networks to improve data rates and system throughputs. This is a brilliant idea for MBS networks, but it will not produce any satisfactory outcome in heterogeneous networks (HetNets).

Studies have inquired into several dynamic power control schemes along with channel assigning methodologies from a game theoretical perspective in [17], [18] but, there is no clear indication whether this could be applicable in HetNets or not. The idea of multi-cell uplink resource allocation together with successive group decoding is formulated as a joint channel scheme with a view to reducing uplink co-tier interference [19]. There is a strong probability of this concept being not functional for reducing cross-tier interference in HetNets. Based on variable channel conditions, channel variation schemes can be used in the time and frequency domains by optimizing the resource pattern and introducing a group lasso term, which reduces interference [20]. Researchers have proposed a stochastic approximation algorithm in HetNets for downlink power control based on inherent channel measurement report feedback in macrocell signaling. These studies demonstrate that FBS will have the authorizing power to update its downlink transmission power by eavesdropping on racial feedback signals from contiguous macrocell user equipment (MUE) to the MBS without demanding excessive backhaul signaling from an MBS [21]. When FBS have the authorization of being self-updated with numerous information, cost effective planning along with end to end encryption security management in a variable channel scheme will be quite difficult.

A distributed power control algorithm can be used to address the uplink interference management problem in cognitive radio networks where the secondary users (SUs) impart the identical licensed spectrum; the primary users do this in multi-cell environments [22]. Furthermore, many authors have proposed different types of frequency allocation schemes, including static and dynamic frequency reuse, cell sectoring, and fractional frequency reuse concept to reduce the effect of interference and increase the spectral efficiency of the integrated network [23]–[25]. Limitation occurs in different cases of these propositions when an economic issue arises. In addition to this, some proposals may fail to ensure better QoS along with efficient femtocell access strategies while reducing interference. Recently, different types of access control schemes such as open access, closed access, and hybrid access have become popular and dependable for their superior performance in interference management. Researchers have proposed that the open access mode is preferred by industrial users, whereas the closed access mode is recommended for the residential users [26]. In the closed access mode, specific user equipment (SUE), i.e., subscribers have the privilege of accessing the FBS owing to the improvement in their own system throughputs and network coverage. However, nonsubscribers are unable to penetrate the closed access FBS. Some researchers have suggested the shared access policy to be used in order to solve this conflict [27]. Although this scheme can reduce interference, no satisfactory outcome appears in the case of reduction of handover and QoS improvement. Furthermore, numerous researchers have taken into account the economic concerns of QoS and brought forward a gaming algorithm for modeling different access strategies to be adopted in HetNets [28]. In addition, many researchers have suggested a hybrid access policy to be used for reducing co-channel interference to allow a limited number of nonsubscribers to connect to the FBS and obtain open access scheme in order to increase the average throughput of the MUEs [29], [30]. If the FBSs are operated in the closed access mode and use the same frequency spectrum based on the MBS, the game theoretical approach can be adopted to reduce the strong interference that FBSs will cause to the user equipment (UE) situated close to the FBSs. Prioritized access issue should be used to make the system more feasible. Interference from FUEs to MBSs is high in the closed access mode in comparison with the open access mode. The hybrid access policy allows nonsubscribers to provide limited connections to FBS [31], [32]. Addition of dynamic channel allocation along with prioritized access will make the system more stable.

Some authors have taken the homogeneous spatial point process into account for uplink capacity analysis and suggested this scheme as a suitable interference avoidance strategy for the HetNets [29]. However, this concept may prove to be ineffective for downlink capacity analysis for the same time being as the concept is based on a single parameter. The open access mode can be a good choice to reduce the effect of interference and improve system throughput for the entire network under a feasible number of FBSs and UEs [33] but, the concept fails when there is large number of UEs. Furthermore, some researchers have described the cooperation between different UEs and proposed a cooperative power game. Using this power game algorithm, they have illustrated the existence of Nash equilibrium (NE), which can be accepted as a stable solution to deal with the access control problem and the interference problem in HetNets [26]. Undoubtedly, adopting a variant access policy for different aspects is not so beneficial in the present as well as future generation wireless networks. Therefore, we cannot propose these schemes for the next generation networks because it will increase the cost to the consumers. Light-emitting diode (LED)-based visible light communication is an emerging trend in the evolution of wireless communication. Some researchers have proposed indoor optical femtocells to be used in lieu of conventional femtocells in order to deal with the challenges of interference reduction and improvement of channel utilization [34]–[35]. While ensuring efficient access control mechanism along with economy considerations, it will be a great challenge to implement optical femtocells in HetNets without proper channel allocation and dynamic power control strategy. We can surely consider this phenomenon as a limitation for the corresponding work. Our main contributions in this paper can be listed as:

(i) We propose an advanced hybrid access policy to be adopted along with prioritized access strategy where there will be no fixed accessing scheme for a particular UE. FBS access will depend on the assigned priority of the UEs. We propose that the corresponding cells possess the property of self-optimizing power control where prioritized access policy will be used. Priority-based access scheme provides better performance in different cases as compared with the conventional systems described in our previous study [36].

(ii) For the channel assigning strategy, we had earlier considered the allocation to be done dynamically on the basis of the cell selection game [37]. Now, we also consider the concept of primary-SUs and propose a dynamic channel allocation scheme based on the game theory.

(iii) We propose the implementation of LED-based femtocells along with conventional femtocells where the typical lighting control function is dimming control and the channel reuse concept is applicable. It is easier to implement an optical femtocell network for the residential users because LEDs are adopted there for general illumination. There will be less probability of the FBSs causing severe interference as there is prioritized access scheme along with dynamic power control strategy is used.

(iv) We investigated the existence of the NE condition and justified our model to satisfy this condition. Our proposed scheme provides a better SINR level, reduces loopholes, and improves system throughput capacity; it also provides better revenue for operators in comparison with the other works while ensuring optimal prices for consumers.

(v) From the thousands of simulation results, we have justified our statements, which are clearly illustrated in Section III with proper explanations. The total work is a combination of QoS provisioning in HetNets along with reduced interference and improved capacity. We have also taken care about the economic perspective and ensure a good revenue for the operators while confirming a reasonable service cost for the consumers.

The remainder of this paper is organized as follows. In Section II, the general architecture of the system model is presented. Mathematical modeling for cases in different scenarios is also explained in this section. Performance evaluations of this paper are illustrated in Section III with the required figures and explanations. Finally, we have summarized our study in Section IV.

## II. SYSTEM MODELLING

In this research, we divided our works into different subsections. At first, we concentrated on Okumura Hata model and prepared a free space propagation model. The output of this model is illustrated in performance evaluation section where the necessity of femtocell in improving QoS and reducing loss levels have been discussed. It is very important to model proper access control strategy in a HetNet otherwise adequate interference will deteriorates system performance. We have prepared an efficient femtocell access strategy on the basis of game theoretical algorithm. In [29], [33] there are conflicts between where to use hybrid access or open access. We suggest to use a newly designed hybrid access policy where there will be prioritized access scheme and the channels to be allocated dynamically that clears the limitations of [14], [21], and [22] for power control strategy. Channel allocation algorithm as well as SINR calculation model are illustrated in the following two subsections. Our model can provide better QoS, offer better capacity with a reduced cost and the operators will earn a good revenue as well. We have to use different notations throughout the modelling. The major nomenclatures used throughout the paper is listed in Table I.

### A. Free Space Propagation Model

Nowadays, interference management along with ensuring better QoS in a cost-effective genre is a serious concern in heterogeneous networks. In addition, penetration loss, shadowing deviation, and free space propagation loss play vital roles in affecting the system performance in numerous ways. They may reduce the system capacity of the wireless network, increase the outage probability of a particular or random user, increase the noise figure, create impediments while selecting the required cell, and so on. [13]. For a macrocellular network, we can consider the generalized free space propagation model with some modifications that can be expressed as follows [10], [13]:

In the indoor environment,

$$L_{\text{Macrocell}} = 36.55 + 26.16\log_{10} f_{m,tr} - 13.82\log_{10} h_{base} - A_M \\ + [44.9 - 6.55\log_{10} h_{base}]\log_{10} d + L_{sh} + L_{pen} \quad [dB] \quad (1)$$

where $A_M$ is defined as follows:

$$A_M = 10.24[\log_{10}(11.75 h_{ms})]^2 - 4.97, \quad 200 \le f_{m,tr} \le 1500 \quad (2)$$

In (1), (2), $f_{m,tr}$ is used as the center frequency (for transmission) of the macrocell in MHz; $h_{base}$ and $h_{ms}$ are defined as the height of the MBS and mobile station (MS) in meters, respectively; $d$ is the distance between the MBS and the MS in kilometers, $L_{sh}$ and $L_{pen}$ are the shadowing standard deviation and the penetration loss, respectively. $L_{pen} = 0$ for outdoor microcell users. Let us assume that all the users receive signals from the FBS that is located on the outside of the indoor arena. Then, free space propagation for femtocell (outdoor user) can be expressed as [13]:

$$L_{femtocell} = 20\log_{10} f_{f,tr} + N\log_{10} d_f + 4.4n^2 - 28 \; [dB] \quad (3)$$

where, $f_{f,tr}$ is the transmission frequency of the femtocell in MHz; $n$ is the number of walls between the MS and the FAP; and $d_f$ is the distance between the FAP and the MS in meters.

### B. Access Strategies

In this subsection, we propose efficient and effective access strategies for the consumers, these strategies are to be adopted to access the femtocell or the microcell. The concept of cognitive radio is used here to model and analyze the attributes of spectrum sharing and interference control between the femtocell and macrocell. In a heterogeneous network, we may observe a co-tier or cross-tier interference. Possible interference scenarios are illustrated in Fig. 1. Generally, we observe a co-tier interference when there is FUE interference to the neighboring FBS in the uplink. Femtocells can be deployed very close to each other inside the apartments or outside as required, in an arbitrary manner. It is possible that the wall separation is inadequate to avoid causing interference to each other. The overall interference can be higher in the case of densely deployed femtocell areas because there are a number of interferers. If the SINR at a certain location (FBS area) is lower than a defined threshold value owing to co-tier interference, it would not be possible to establish a communication link; thus, a dead zone would be created. Any UE in the neighboring femtocells, which has high-power transmission, will affect the victim femtocell and the system performance will be degraded. Apart from this, we can experience cross-tier interference when FBS causes interference to the downlink of MUE; also, MUEs can cause interference at the uplink of a nearby FBS. When MUEs receive rigid signals from the neighboring cell to which access is denied, then there would be a possibility of dead zones occurring around the femtocell. The area around the femtocell becomes a dead zone for MUE if the closed access mode is chosen. Furthermore, there is a high probability of power leakage through windows, doors, ventilators, and so on because of close deployment of femtocells. Considering, all these cases we strongly believe that dynamic spectrum allocation is required along with efficient femtocell access strategy.

TABLE I
NOMENCLATURE

| Parameter | Value |
|---|---|
| *Sub-section II A* | |
| $f_{m,tr}$ | Center frequency of Macrocell for Transmission |
| $f_{f,tr}$ | Center frequency of Femtocell for Transmission |
| $h_{base}$ | Height of Base Station |
| $h_{ms}$ | Height of Mobile Station |
| $d$ | Distance between the MBS and the MS in kilometers |
| $d_f$ | distance between the FAP and the MS in meters |
| *Sub-section II B* | |
| $S_i$ | complete set of strategies with player $i$ |
| $P$ | Finite set of players |
| $P_{sub}$ | Number of Players who are allowed to access FBS (Subscribers) |
| $P_{non}$ | Number of Players who are not allowed to access FBS (Nonsubscribers) |
| $p$ | Number of players who are not connecting to MBS |
| $U_i$ | Utility Function based on the player strategies |
| $\Phi(S)$ | Potential Game Function |
| $(AP)_{max}$ | Maximum allowable power of FBS for data transmission purposes during time slots |
| $(TP)_{MAX,k}$ | Maximum transmission power of BS "$k$" |
| $L_{m,k}^{MUE}$ | Large-scale channel gain between BS and MUE. |
| $B$ | Usable bandwidth of the system, |
| $L_{\hat{p}}$ | Large-scale channel gain |
| $\xi^2 P$ | Additive white Gaussian noise power |
| $\chi$ | Data speed limit |
| $\phi$ | Price that the consumer should pay to the service providing operators |
| $\Delta$ | Periodic adjustor |
| *Sub-section II D* | |
| $\Re$ | Channel realization function to be used for MBS, FBS/FAP and OFAP |
| $P_{k/j/m}^{(n)}$ | Power Allocation Vector |
| $\gamma$ | SINR Level |
| *Sub-section II E* | |
| $\varpi$ | Threshold value of SINR |

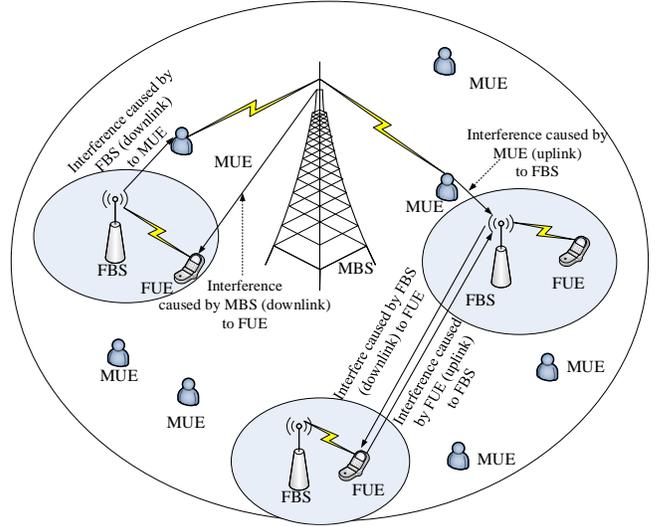

Fig. 1. Interference scenarios in heterogeneous networks.

The prime motive of the licensed spectrum is to permit the UEs to access the microcell or femtocell as per requirements. Meanwhile, FBSs are taken into account as subsidiary ingredients to reduce the probability of spectrum reuse. When a particular channel is brought to service by an MUE, FUEs are capable of transmitting data only in the channel. A typical outlook of a HetNet scenario along with different types of UE are illustrated in Fig. 2, where FBS is enabled with required lighting arrangements (marked red or green). Let $D$ and $D^*$ be defined as two sets of UEs for which reference signal receiving power (RP) from FBS is comprehensive and more diminished than that from MBS. According to the proposition, all subscribers are considered to be in the set $D$; therefore, they can access the subscribing femtocell. The nonsubscribers are classified based on their availability in $D$ or $D^*$ sets. Generally, the nonsubscribers, who are in the $D^*$ set, possess the aptitude to select a macrocell as their intended network. However, to connect and explore maximum utilities, nonsubscribers who are in the $D$ set, have the option of choosing between FBS or MBS. The set of nonsubscribers $\in D$ are considered as the players in the cell selection game. It is assumed that the subscribers are authorized to be connected to the FBS only. The subscribers have the ability to affect the game by providing adequate interference. Both the nonsubscribers $\in D^*$ and the macro players are considered as MUEs whereas the subscribers and femto players are considered as FUEs. According to the fundamentals of game theory, the players are treated as the decision makers in a game where various types of actions may be incorporated as attainable options. The players of a game are capable of picking up their own strategies and the resulting strategy profiles settle the outcome of the game. All the possible yields of players are evident from the utility function. Modulation schemes, transmit power levels, and so on are considered here as a set of actions. Different performance metrics, i.e., system throughput, SINR capacity, and so on are assumed to be a set of preferences for this gaming model.

Generally, channel allocation in a networking game can be modeled mathematically as follows:

$$C_{Allocation} = \{P, \{S_i\}_{i \in P}; \{U_i\}_{i \in P}\} \quad (4)$$

where $P$ is defined as a finite set of players, and $S_i$ is the complete set of strategies with player $i$. The utility function $U_i$ is a function of $s_i$ (which is the strategy used by the player $i$), and $s_{-i}$ is the strategy profile of its opponent.

If $S = [s_1, s_2, s_3, \ldots, s_N]$ is a set of player strategies, then NE will be fulfilled only if,

$$U_i(S) \geq U_i(s'_i, s_{-i}), \quad \forall i \in N, s'_i \in s_i \quad (5)$$

Let us consider a heterogeneous network where a cell selection game is defined as follows [29]:

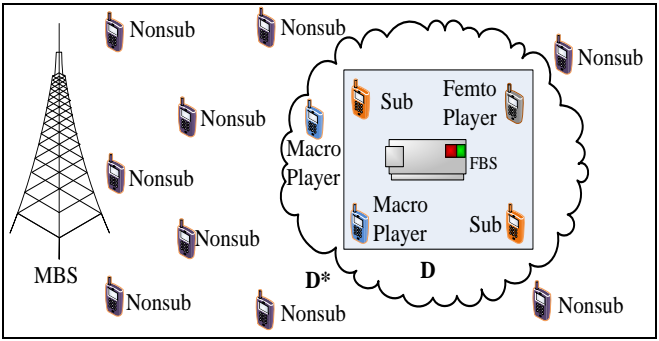

Fig. 2. Layout of a heterogeneous network with players' definition.

$$\langle Y, (S_i)_{i \in 1}, (u_i)_{i \in 1} \rangle \quad (6)$$

where, $Y = \{1, 2, 3,..., N\}$ is the set of nonsubscribers $\in P$, which are assumed as the finite set of players. Let us define "1" and "0" as pure strategy for connecting to MBS and FBS, respectively, for each player. We can define $S = \prod_i S_i$ as the set of action profiles. Now, the utility function for each player $i$ can be formulated as follows:

$$u_i(s_i, s_{-i}) = \begin{cases} u_1(1, \|s_{-i}\|_0); & s_i = 1 \\ u_0(0, \|s_{-i}\|_0); & s_i = 0 \end{cases} \quad (7)$$

where $u_0$ and $u_1$ represent the utility functions of the players connecting to FBS and MBS, respectively. The strategies can represent different kinds of action profiles. For $N = 6$ and $\|s_{-i}\|_0 = 3$, we obtain the action profiles as follows: $(1, (1,1,1,0,0)), (1,(1,1,0,1,0)), (1, (1,1,0,0,1)) \ldots \ldots \ldots i.e., u_i(1,(1,1,1,0,0))$
$= u_i(1,(1,1,1,0,0)) = u_i(1,(1,1,0,1,0)) = u_i(1,(1,1,0,0,1)) = \cdots \cdots = u_i(1,3)$

Cell selection game defined in (6) is also a potential game and has a pure strategy NE. According to the game theoretical algorithm and potential game formulation, it can be formulated as follows [29]:

$$\Phi(s) = \tilde{\Phi}\|s\|_0 = \sum_{n=0}^{\|s\|_0} \tilde{u}_0(n) + \sum_{n=\|s\|_0}^{P-1} \tilde{u}_1(n) \quad (8)$$

Existence of NE for this definition is illustrated in [29], and this confirms the proof. In the round-robin scheduling scheme for the downlink scenario, the time frame is partitioned into equal slots. Required scheduling in macrocells is done with $P_{non}$ nonsubscribers $\in D^*$, and $p$ players connecting to FBSs with $P-p$ players connecting to MBS. This scenario clearly implies that there are total $P_{sub} + P - p$ MUEs connecting to MBS. For each UE, the RP from the serving cell and the neighboring cell are estimated by UE. Let the RP of the $m$th FUE received from the BS $k$ ($k = 1$ for MBS, and $k = 0$ for FBS) be written as follows [37]:

$$(RP)_{m,k}^{MUE} = (TP)_{MAX,k} \cdot L_{m,k}^{MUE} \quad (9)$$

where $(TP)_{MAX,k}$ is defined as the maximum transmission power of BS "$k$", and $L_{m,k}^{MUE}$ represents the large-scale channel gain between BS and MUE.

Let the maximum allowable power of FBS for data transmission purposes during time slots $T_m$ be denoted as $(AP)_{max}$; it is defined as follows [37]:

$$(AP)_{\max} = \begin{cases} (TP)_{MAX,\,0}, & (RP)_{m,1}^{MUE}[dBm] \geq Q_{m,1}^{MUE}[dBm] + \delta + \Omega \\ 0, & (RP)_{j,1}^{MUE}[dBm] < Q_{j,1}^{MUE}[dBm] + \delta + \Omega \end{cases} \quad (10)$$

where the parameters $\delta$ and $\Omega$ are defined as the threshold levels to determine the co-channel, adjacent channel interference and the co-tier, cross-tier interference level, respectively.

Here, the allowable power is estimated for the worst player. Considering $m$ players in action, the cell capacity of FBS is formulated as follows [35]:

$$C_{FBS}^{\hat{p}}(m) = \frac{B}{P_{non} + P - p} \sum_{m=0}^{P_{non}+P-p} \log_2\left(1 + \frac{(AP)_{\max} \cdot L_{\hat{p},0}}{\xi^2 P + (TP)_{MAX,1} \cdot L_{\hat{p},1}}\right) \quad (11)$$

where $B$, $L_{\hat{p}}$, and $\xi^2 P$ denote the usable bandwidth of the system, large-scale channel gain and additive white Gaussian noise power, respectively. FBS has no authorization to transmit any power when macro players keep common slots in operational activities. During such cases, the FBS will provide robust interference to macro players if the FBS transmits during $T_m$ because $(RP)_{m,0}^{MUE} + \delta + \Omega > (RP)_{m,1}^{MUE}$; this condition is valid if it satisfies the condition $P_{non} + P - p \geq m \geq P_{non} + 1$. On the basis of the true-false concepts of the fundamental game theory [27], the modified formulation of (11) is as follows:

$$C_{FBS}^{\hat{p}}(m) = \frac{(P_{non} - Z)B}{P_{non} + P - p} \log_2\left(1 + \frac{(TP)_{MAX,0} \cdot L_{\hat{p},0}}{\xi^2 P + (TP)_{MAX,1} \cdot L_{\hat{p},1}}\right) \quad (12)$$

Subscribers are assumed to possess the authorized privileges to access the FBS; therefore, the concept of assigning dynamic priority for primary and SUs arises. According to [29] and [37], we propose a system control parameter $\beta$ to be used in the modification of (8). This proposition indicates that all subscribers can firstly be allocated the ratio $\beta$ of total bandwidth available for the femtocell; thereafter, all the FUEs will share the remaining $(1-\beta)$ resource. For $m$ players, subscribers are allowed to share $\frac{\beta P + P_{sub}}{P_{sub}^2 + P_{sub} \cdot P}$ of the available resource. FBSs are authorized to control the total number of its serving UEs by considering the value of the closed rate $\beta$. In our work, we have discussed the effects of $\beta$ in different access modes i.e., open, close, hybrid, etc. and illustrated the effects graphically in Figs. 10-13 which proves the superiority of our propositions compared to the proposals in [29] and [33]. $\beta$ is required to calculate the utility function which has an immense effect in deciding the attributes of the UEs. Thereafter, the final utility function $u_i(s)$ for each player $i$ can be formulated as follows:

$$u_i(s) = \begin{cases} u_1(\|s\|_1) = \dfrac{B}{P + P_{non} - \|s\|_0} \log_2\left(1 + \dfrac{P_{MAX,0} \cdot L_{\hat{p},1}}{\zeta^2 P}\right), & s_i = 1; \\ u_0(\|s\|_0) = \dfrac{1-\beta}{P_{sub} + \|s\|_0} \cdot C_{FBS}^{\hat{p}}(\|s\|_0) - \chi \cdot \phi \cdot \Delta; & s_i = 0. \end{cases} \quad (13)$$

where $\chi$ is the data speed limit, $\phi$ is the price that the consumer should pay for the service-providing operators, and $\Delta$ is a periodic adjustor that can be adjusted dynamically as per requirements with a precautionary notice to the consumers. Therefore, we can define $\chi \cdot \phi \cdot \Delta$ as the revenue that the operator can attain in this scenario. Each femto player can be allocated $\left(\dfrac{1-\beta}{P_{non} + P - p}\right)$ of the available resource. We can observe the superiority of our model for economic perspective in performance evaluation section.

### C. Dynamic Channel Allocation Scheme

An efficient and cost-effective channel allocation scheme is proposed here on the basis of the game theoretical framework. MBS settles and organizes its power level for different sets of users and allocates channels for both FBS and optical femto access points (OFAPs). FBS and OFAP always collect feedback information from their environment and employ this information to allocate the required channel to the voice call users and data users. Furthermore, the collected information assists in alternating between the required strategy to be adopted by FBSs and OFAPs dynamically with a view to providing better QoS. In the proposed scheme, first, a definite number of channels are allocated for both FBS and OFAP. Here, femtocells and optical femtocells are assigned for voice call users and data users, respectively. Telecom service-providing operators will assign required channels for FAPs and OFAPs considering the user-handling capability. Generally, these channels have tight constraints of maintaining a fixed number of users. It is possible that the channel has some unused space but no users are acquiring it.

The operators impose the cost for these unused spaces also on the consumers. The consumers have to pay for it because they are charged a flat tariff system. We have considered the channel tunability concept where the unused spaces of a particular channel can be used by the users of another channel if required. FBS and OFAP always keep themselves engaged in counting the number of users under their provision. When FBS or OFAP find any unused space within their capacity limit and receive space-acquiring requests from the other side, then FBS and OFAP allocate the unused number of channels to the required data users and voice call users, respectively. In this case, they control the lighting arrangements (shown in Fig. 3) automatically and assign space for the required purpose. The reason behind this proposition is as follows. If there is no data user but the lighting arrangements are active, then there will be unused bandwidths because of higher bandwidth transferability of the light waves as compared to the radio waves. In this case, the consumers will have to bear the additional charges even though they have not used the services. We can confidently assert that there is no need to activate the lighting arrangements in the femtocells for the voice call users only. The cells will be designed according to the gaming algorithm, and they will activate lighting arrangements just when there are requests from the data users. In this way, the consumers will have to pay reduced costs and the operators can use their unused bandwidths properly as per requirements. The summary of the total algorithm is illustrated in a flowchart in Fig. 3. We propose that this channel allocation and the lighting arrangements are to happen dynamically. We have successfully justified our proposition by numerous simulation results illustrated in Section III.

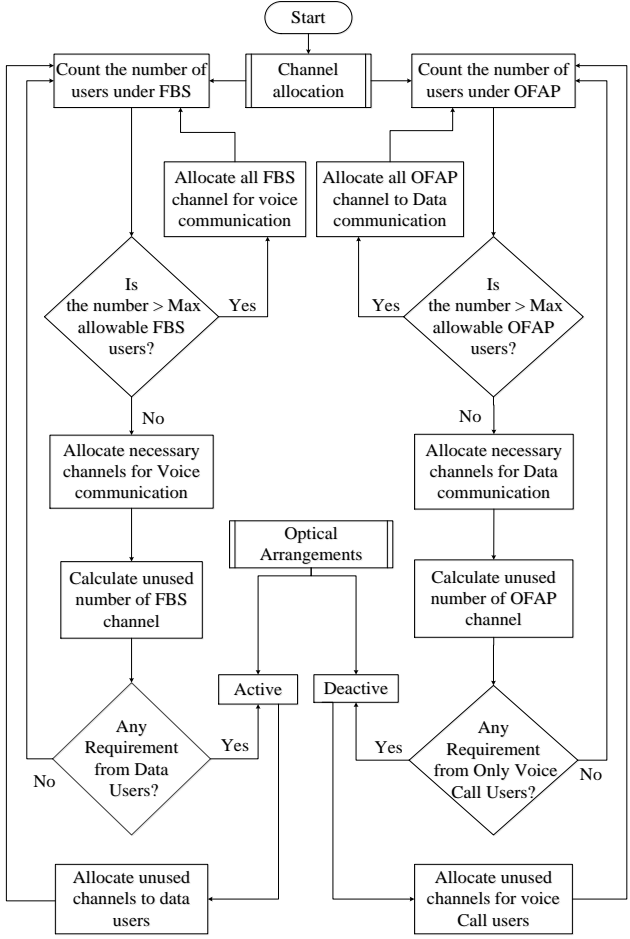

Fig. 3. Proposed channel allocation scheme.

TABLE II
POSSIBLE COMBINATION OF CHANNEL REALIZATION

| Relation | Channel Realization | Between |
|---|---|---|
| $a,b,c,j,k,m \in H^2 \times N$ | $\Re_{1,m,c}^{(n)}$ | MBS m & MUE c |
| $a,b,c,j,k,m \in F \times H \times N$ | $\Re_{2,k,c}^{(n)}$ | FAP k & MUE c |
| $a,b,c,j,k,m \in O \times H \times N$ | $\Re_{3,j,c}^{(n)}$ | OFAP j & MUE c |
| $a,b,c,j,k,m \in F^2 \times N$ | $\Re_{4,k,c}^{(n)}$ | FAP k & FUE b |
| $a,b,c,j,k,m \in H \times F \times N$ | $\Re_{5,m,c}^{(n)}$ | MBS m & FUE b |
| $a,b,c,j,k,m \in O \times F \times N$ | $\Re_{6,j,c}^{(n)}$ | OFAP j & FUE b |
| $a,b,c,j,k,m \in O^2 \times N$ | $\Re_{7,j,c}^{(n)}$ | OFAP j & OFUE a |
| $a,b,c,j,k,m \in H \times O \times N$ | $\Re_{8,m,c}^{(n)}$ | MBS m & OFUE a |
| $a,b,c,j,k,m \in F \times O \times N$ | $\Re_{9,m,c}^{(n)}$ | FAP k & OFUE a |

*D. Analytical Model for SINR Calculation*

In this subsection, we have explained the mathematics behind our proposed model on the basis of the gaming approach along with the required access scheme. Let us consider a set H = {1, 2, 3,…, m} of MBS where each of the elements is supposed to operate over a monopolistic frequency band. Each MBS follows a time division multiple access (TDMA) strategy to remain active in the service of the MUEs. It is assumed that there exists a set of N = {1, 2, 3, …, n} frequency bands such that MBS can maintain its operational activities in a proper way without any disturbance. A set of F = {1, 2, 3, …, f} of FAPs and O = {1, 2, 3, …, q} of OFAPs have been considered for preparing a mathematical model of HetNets. Following a TDMA policy, each femtocell and optical femtocell is allowed to use any of the available frequency bands with a view to providing services to the corresponding FUEs and OFUEs [25], [29], [36]. There can be various combinations of channel realizations, which are listed in Table II. All assumptions are considered for a discrete time index, which is defined as $t \in$ {1, 2, 3, …, ∞}. The vector of all channel realizations is denoted by $\Re(t)$ at the discrete time index t, and each component of $\Re(t)$ is identically distributed following a probability distribution. Let the N-dimensional vector,

$$P_k(t) = \left[P_k^{(1)}(t), P_k^{(2)}(t), P_k^{(3)}(t) \cdots P_k^{(n)}(t)\right]; P_m(t) = \left[P_m^{(1)}(t), P_m^{(2)}(t), P_m^{(3)}(t) \cdots P_m^{(n)}(t)\right]; \text{and } P_j(t) = \left[P_j^{(1)}(t), P_j^{(2)}(t), P_j^{(3)}(t) \cdots P_j^{(n)}(t)\right]$$

denote the power allocation vector of FAP $k \in F$, MUE $m \in H$ and OFAP, $j \in O$, respectively, for the time index *t*. Here, we used these power allocation vectors and the various combinations of channel realization listed in Table II with a view to calculating SINR levels for the FBS, MBS and OFAP, respectively.

For all $k \in F$, $m \in H$, and $j \in O$, we can assume (14), (15), and (16) in order to determine the SINR levels of FUE, MUE, and OFUE where, $\gamma_k^{(n)}$, $\gamma_m^{(n)}$, and $\gamma_j^{(n)}$ are defined as the SINR of FUE *k*, MUE *m*, and OFUE *j*, respectively.

$$\gamma_k^{(n)}(t) = \frac{P_k^{(n)}(t)\left|\Re_{4,k,b}^{(n)}(t)\right|^2}{\left(\left(\sigma_k^{(n)}\right)^2 + \prod_{j \in O} P_j^{(n)}(t)\left|\Re_{6,j,b}(t)\right|^2 + \prod_{m \in M} P_m^{(n)}\left|\Re_{5,m,b}(t)\right|^2 + \prod_{i \in F \setminus k} P_i^{(n)}(t)\left|\Re_{4,k,b}^{(n)}(t)\right|^2\right)} \quad (14)$$

$$\gamma_m^{(n)}(t) = \frac{P_m^{(n)}(t)\left|\Re_{1,m,c}^{(n)}(t)\right|^2}{\left(\left(\sigma_m^{(n)}\right)^2 + \prod_{k \in F} P_k^{(n)}(t)\left|\Re_{2,k,c}(t)\right|^2 + \prod_{m \in M} P_m^{(n)}\left|\Re_{3,j,c}(t)\right|^2 + \prod_{r \in M \setminus m} P_r^{(n)}(t)\left|\Re_{1,m,c}^{(n)}(t)\right|^2\right)} \quad (15)$$

$$\gamma_j^{(n)}(t) = \frac{P_j^{(n)}(t)\left|\Re_{7,j,a}^{(n)}(t)\right|^2}{\left(\left(\sigma_j^{(n)}\right)^2 + \prod_{m \in M} P_m^{(n)}(t)\left|\Re_{8,m,a}(t)\right|^2 + \prod_{k \in F} P_k^{(n)}\left|\Re_{9,k,a}(t)\right|^2 + \prod_{w \in O \setminus j} P_w^{(n)}(t)\left|\Re_{7,j,a}^{(n)}(t)\right|^2\right)} \quad (16)$$

In (14)-(16), $\left(\sigma_k^{(n)}\right)^2$, $\left(\sigma_m^{(n)}\right)^2$, and $\left(\sigma_j^{(n)}\right)^2$ are assumed to be noise power values over MUE, FUE, and OFUE, respectively. All FBSs are inquisitive of the optimizing interference, and the users cooperate to ensure a fair allocation.

*E. Outage Probability Analysis*

Considering all the interfering neighboring macrocells and femtocells, we can evaluate the outage probability, which can be formulated as follows [36]:

$$P_{outage} = 1 - e^{\left(-\frac{\varpi}{SINR}\right)} \quad (17)$$

where $\varpi$ is defined as the threshold value of SINR and $SINR < \varpi$. The values of *SINR* can be calculated from (14), (15), and (16) as per requirements.

III. PERFORMANCE EVALUATION

In this section, we justified the effectiveness of our proposed system model. Let us consider a HetNet, as illustrated in Fig. 3, where there is one MBS with transmission range radius equal to 1500 m and one FBS located in the center of the room having dimensions of 25 m × 25 m. In order to run a simulation of our proposed model, we took into account the large-scale channel model, which included path loss, shadowing, and wall penetration loss. We fixed the locations of indoor and outdoor UEs considering the uniform distribution property within the room and the maximum transmission range of MBS. With an increase in population, there is a severe paucity of open ground space in the mega cities. This has led to the establishment of densely populated femtocells in residential buildings. A common scenario nowadays in this type of highly populated large building areas is that the users who reside in high-rise 912buildings or indoor environments right behind tall buildings are deprived of adequate network coverage. For such situations, we have analyzed the performance of a HetNet in Figs. 4–7. The users living in this kind of environment receive low signal levels because of high loss levels. We can see the effect of formation of loopholes because of congestion and penetration losses of the building walls in Figs. 4 and 6, respectively. As a result, the signal cannot penetrate and the users receive poor network coverages. It may hamper both the voice call users as well as the data users because of low signal power levels. The users will also experience poor service when many UEs will take attempt to make calls or browse internet. Implementation of the femtocell in the congested areas can eliminate the loopholes and provide better signal levels to the UEs, which offer low loss levels, thereby ensuring better signal power levels which is illustrated clearly in Figs. 5 and 7. Femtocell has the unique characteristics of being capable to work as small power BS which has led it to be used in this type of scenarios. We have provided 3-D plots to figure out the effect of our proposal and make our propositions more feasible. Both the consumers as well as the operators will be greatly benefitted with the improved coverage and signal strength. As soon as a mobile phone detects femtocell is ready to use, it will start to consume less power with a view to communicating with it and thus increases the battery life also. Femtocell installing position plays a vital role as it may happen that, installed femtocell is using the same frequency bands dedicated for microcell which will create severe interference. We proposed adaptive power control scheme to be used and also modeled interference mitigation scheme which is discussed earlier in section II and we will observe the performance evaluation of that propositions in the next figures.

TABLE III
SYSTEM PARAMETERS

| Parameter | Value |
|---|---|
| Carrier frequency | 1800 [MHz] |
| Transmitted signal power by the MBS | 1.5 [kW] |
| Transmitted signal power by a FBS | 15 [mW] |
| Height of a MBS | 75 m |
| Height of a FBS | 3 m |
| Number of sub-channels | 30 |
| System bandwidth | 5.5 [MHz] |
| Number of indoor *subscribers* | 5 |
| Number of indoor *nonsubscribers* | 8 |
| Number of outdoor *nonsubscribers* | 10 |
| Noise figure | 9 [dB] |
| Channel Noise density | -175[dBm/Hz] |
| Shadowing deviation (MBS → outdoor UE) | 6[ dB] |
| Shadowing deviation (MBS → indoor UE) | 8 [dB] |
| Shadowing deviation (FBS → outdoor UE) | 8 [dB] |
| Shadowing deviation (FBS → indoor UE) | 3 [dB] |
| Free Space Propagation loss for MBS | 10 [dB] |
| Wall penetration loss for MBS | 20 [dB] |
| Wall penetration loss for FBS | 20 [dB] |
| Distance between MBS and FBS | 500 m |
| Threshold value of *SNIR* for MBS | 10 [dB] |
| Threshold value of *SNIR* for outside transceiver | 7 [dB] |

In Figs. 8-9, we have illustrated about the superiority of our works to the conventional MBS [21], FBS [5], [8], and open access OFBS (OFBS) deployments in terms of channel utilization and outage probability performances. A comparison of channel utilization is shown in Fig. 8. It is clear that the percentage channel utilization is very poor in the conventional scheme (MBS deployment) when the number of users is low. It seems that approximately 30% of the channel remains unused, which is a great loss of bandwidth and the consumers have to pay the charges for the service-providing operators as well even though they do not receive services from those unused bandwidths. However, femtocell deployment demonstrates much better channel utilization. Our proposed scheme provides better performance by allocating channels dynamically and dimming control properties of lighting arrangements in OFAP outcomes. The outage probability model is illustrated in Fig. 9. It is seen that call outage probability for MBS deployment cases increases rapidly with the increase of users, whereas it remains almost constant for both FBS and open access OFBS (OFBS) when the number of users is moderate. Nevertheless, it would not be adequate enough in the highly dense femtocell deployment area because there will be a massive chance of being influenced by co-tier and cross-tier interference, which will surely degrade the performance. Our proposed scheme shows the least outage probability, which clearly indicates that this model is more effective than conventional schemes in ensuring QoS.

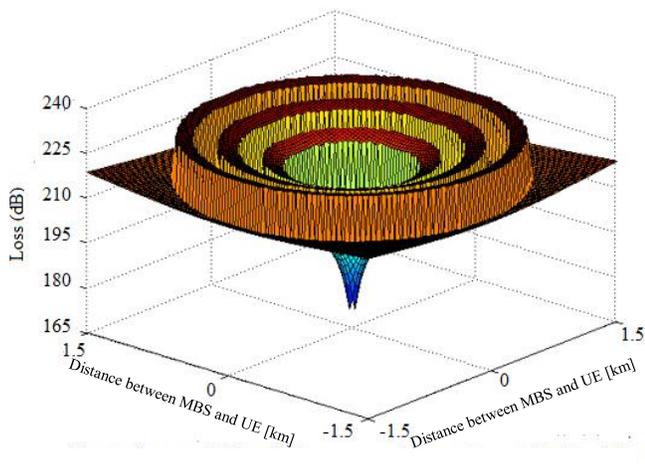

Fig. 4. Loss level (without femtocell)

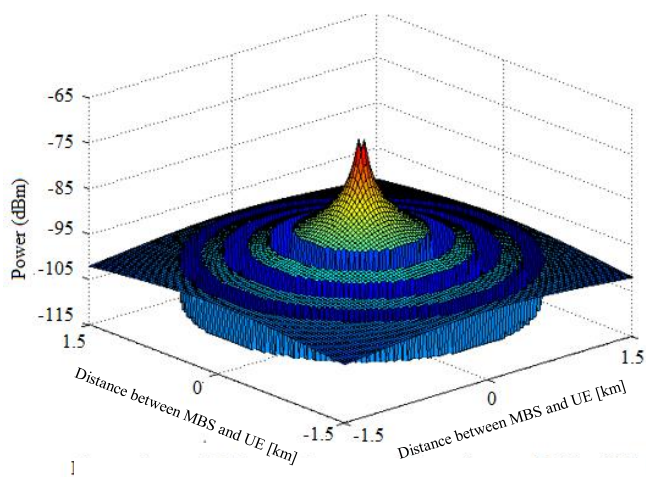

Fig. 6. Received power level (without femtocell)

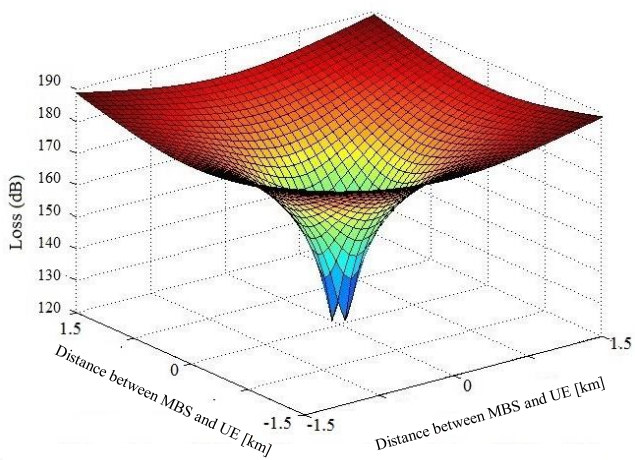

Fig. 5. Loss level (with femtocell)

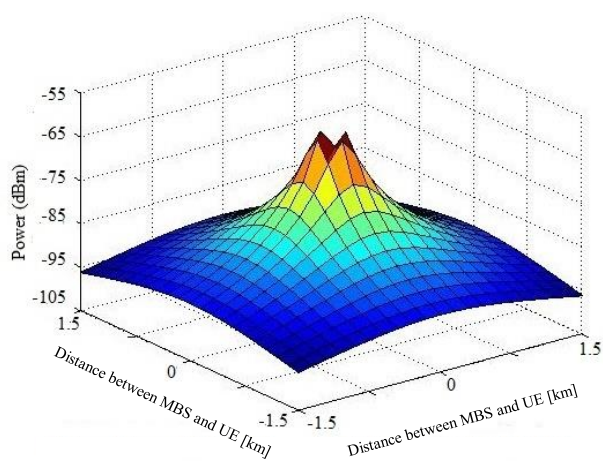

Fig. 7. Received power level (with femtocell)

In Figs. 10-13, we have compared our works with the open access, closed access, and hybrid access FBS deployments [9], [29], [30]. Fig. 10 illustrates the scenario of average capacity of *subscriber* and system UE at NE versus the distance between MBS and FBS in open ($\beta = 0.0$), closed ($\beta = 1.0$), and hybrid (with $\beta = 0.5$) access modes [29] along with our proposed schemes ($\beta = 0.75$). It is clearly illustrated in the figure that, the average capacity of *subscriber* and system increase with the distance as femtocell network is less interfered by MBS when their intermediate distance is quite large. Generally, hybrid access mode can attain better performance for *subscribers* and open access mode is more effective for system UEs [26], [29]. Our proposed scheme can eliminate this contradictory behavior and provide comparatively better balance of performance between subscribers and system UEs compared to open, closed, and conventional access modes. This illustrates the reason behind our strong propositions of the adoptability of this scheme in future generation networks.

Fig. 11 illustrates the cumulative distribution function of capacity at NE for both the subscriber and the system in different access modes. From the subscriber's perspective, better performance can be obtained in the hybrid access mode, whereas the entire system can lead to better performance in the open access mode [26], [29]. After gradually observing an increasing trend of the curve from the open access mode, we can surely conclude that it will be better to adopt the open access mode because spectrum sharing between the UEs can be easily achieved in comparison with other modes. Considering the entire system, it can be intuitively said that the performances of both open and hybrid access modes are better in almost every case than that of the closed access mode. Our proposed scheme can eliminate this contradictory behavior also and produce more satisfactory performances between subscribers and system UEs compared to open, closed, and conventional hybrid access modes. Fig. 12 represents the average capacity of the subscriber and the system UE for NE versus price. From the graphical illustration, it is clear that our proposed scheme is superior to the other schemes because it is beneficial to both the subscriber and the service-providing operators. Our proposed scheme is capable of ensuring better capacity than other schemes for the same price. We have also analyzed the effects of price in the average revenue of the service-providing operators, which is illustrated in Fig. 13. The average revenue in the open access mode is more than that in the hybrid access mode; the difference is quite high. Although the difference in the average revenues between our proposed scheme and open access policy is marginal, our proposed model still ensures better revenue and better capacity. Peak values in Fig. 13 clearly indicate an optimal price that the network operators can charge from the consumers to maximize their revenue. All of these simulation results clearly emphasize the justification for our proposed model, and the operators can adopt our proposed scheme as an optimum solution for operators and consumers because it is beneficial to both the consumers and the operators. Apart from these, another important outcome is that, no network time protocol will be required here for ensuring frequency stability which is being used in general cases. In this way, our proposed model is undoubtedly a cost effective ideal scheme to be adopted by the operators.

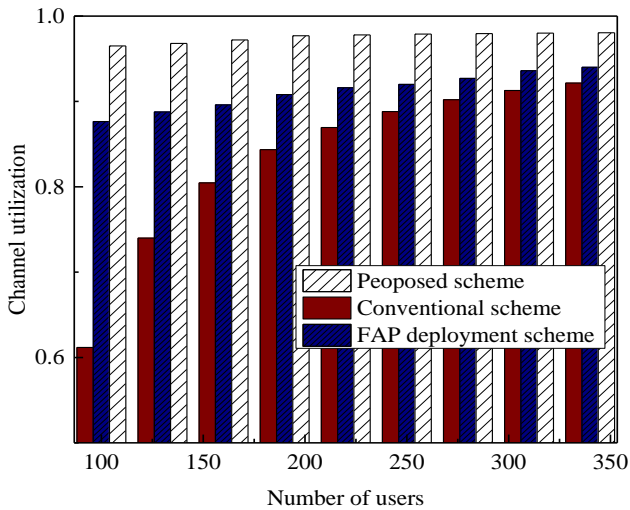

Fig. 8. Channel utilization Comparison in HetNet

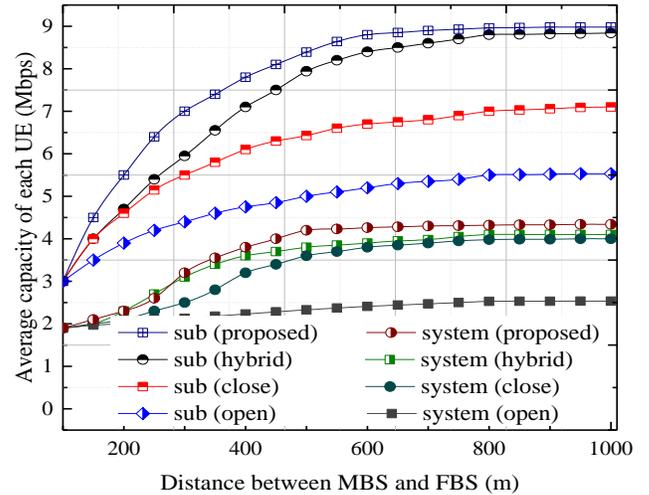

Fig. 10. Average capacity of *subscriber* and system UE at different access modes versus the distance between MBS and FBS.

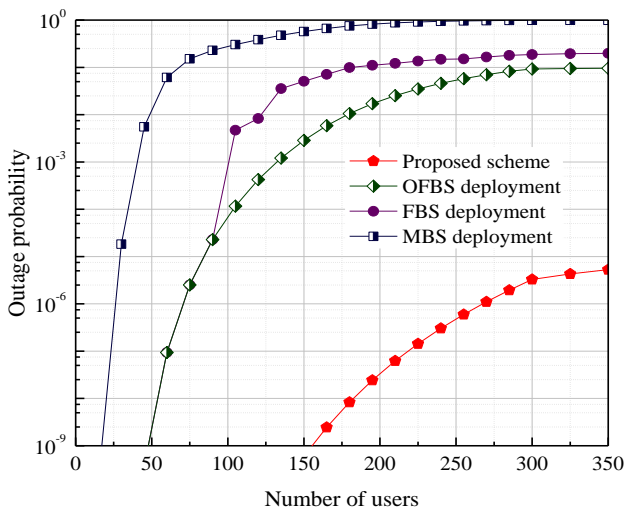

Fig. 9. Outage probability comparison in HetNet

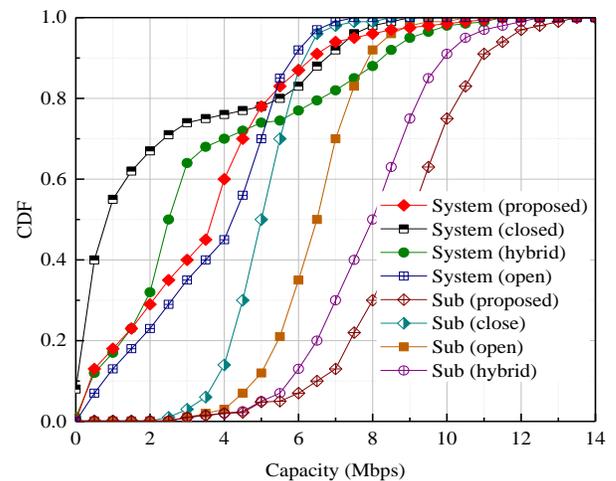

Fig. 11. CDF of capacity at NE for the *subscriber* and system in different access modes.

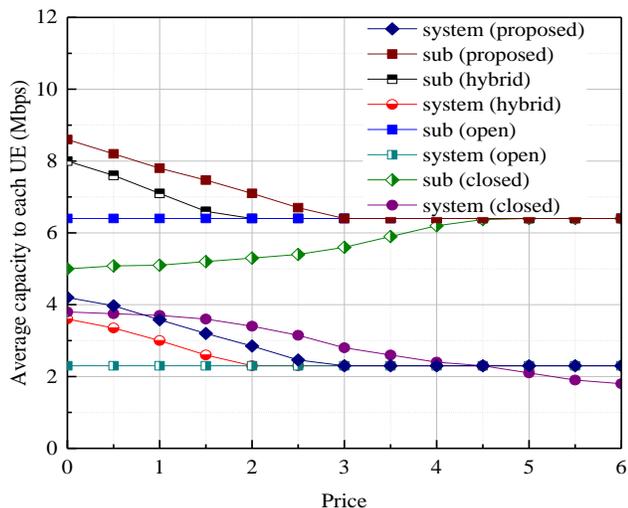

Fig. 12. Average capacity of *subscriber* and system UE at NE versus price.

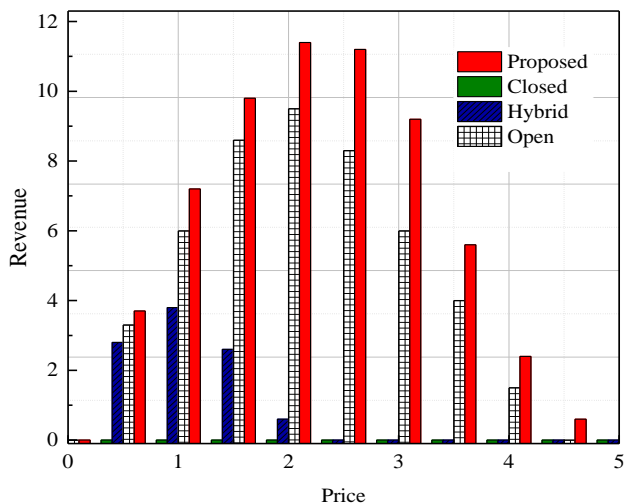

Fig. 13. Average revenue (operator) versus the price.

## IV. Conclusions

In this study, our main objectives were divided into several parts. At first, we proposed the use of femtocells to eliminate the problems in densely populated buildings and to investigate the scenarios after the deployment of femtocells. We considered both the outdoor and indoor environments for justifying our proposed model. Secondly, we worked on interference management strategies in a HetNet because of the unusual cellular operation after the smooth integration of femtocells into the macrocellular network. In this paper, we have proposed an efficient game theoretic algorithm–based mathematical model in which high priority is given to allocating channels in a dynamic genre. Our proposed scheme substantiates low outage probability and better channel utilization and thus ensures better QoS; this clearly indicates that the effect of interference has been reduced. Outage probability depends on the SINR level, which is modeled analytically considering the worst cases. As our scheme provides satisfactory results for the worst cases, we anticipate that our proposed scheme is capable of providing greater flexibility to the entire network. Undoubtedly, it is applicable in the present wireless multi-service networks. Furthermore, we proposed a femtocell access model to reduce the effects of interference on the basis of gaming algorithm wherein we have emphasized the concept of dynamic self-optimizing power control and priority-based access exposure. The existence of NE is investigated, which justifies the stability of our model. Although hybrid access has higher capacity and the open access mode generates better revenue for the operator in comparison to other modes, our proposed scheme can be adopted for increasing flexibility in the performance enhancement of the subscriber, the entire system, and the operators. From the graphical illustrations and discussions stated above (see Section III), it is clear that our proposed scheme can be a good surveyor for the future generation wireless HetNets.